# Anomalous optical response of graphene on hexagonal boron nitride substrates

Adilet N. Toksumakov[1, 2, +], Georgy A. Ermolaev[1, 3, +], Mikhail K. Tatmyshevskiy[1], Yuri A. Klishin[1], Aleksandr S. Slavich[1], Ilya V. Begichev[1], Dusan Stosic[1], Dmitry I. Yakubovsky[1], Dmitry G. Kvashnin[2], Andrey A. Vyshnevyy[1, 3], Aleksey V. Arsenin[1, 3], Valentyn S. Volkov[1, 3,*] and Davit A. Ghazaryan[1,*]

[1]Center for Photonics and 2D Materials, Moscow Institute of Physics and Technology, Dolgoprudny 141701, Russia

[2]Emanuel Institute of Biochemical Physics RAS, Moscow 119334, Russia

[3]Emerging Technologies Research Center, XPANCEO, Dubai Investment Park First, Dubai, United Arab Emirates

[+]These authors contributed equally

[*]Authors to whom correspondence should be addressed: vsv.mipt@gmail.com and dav280892@gmail.com

**Abstract**

Graphene/$h$BN heterostructures can be considered as one of the basic building blocks for the next-generation optoelectronics mostly owing to the record-high electron mobilities. However, currently, the studies of the intrinsic optical properties of graphene are limited to the standard substrates (SiO$_2$/Si, glass, quartz) despite the growing interest in graphene/$h$BN heterostructures. This can be attributed to a challenging task of the determination of $h$BN's strongly anisotropic dielectric tensor in the total optical response. In this study, we overcome this issue through imaging spectroscopic ellipsometry utilizing simultaneous analysis of $h$BN's optical response with and without graphene monolayers. Our technique allowed us to retrieve the optical constants of graphene from graphene/$h$BN heterostructures in a broad spectral range of 250—950 nm. Our results suggest that graphene's absorption on $h$BN may exceed the one of graphene on SiO$_2$/Si by about 60 %.

**Introduction**

Combination of hexagonal boron nitride ($h$BN) with graphene into van der Waals heterostructures attracted much attention at a recent time[1–4]. $h$BN is an insulator with a large bandgap that possesses honeycomb crystal structure commensurate to the one of graphene, but with a slight mismatch of the lattice constants. When assembled into such heterostructures in its high-quality single-crystal form, it provides a suppression of external disorder in graphene and an enhancement of electron mobilities. Thus, it has been proven to be a supreme substrate[5], encapsulating layer[6,7], and tunneling barrier[8,9] in graphene-based electronic devices. Likewise, $h$BN was also found to be an irreplaceable constituent in graphene-based optoelectronic devices, such as photodetectors[10,11], DUV electroluminescent devices[12], THz optoelectronic elements[13,14], and even light bulbs[15]. From the standpoint of optical properties, it is known that the integration of $h$BN with graphene may boost an infrared spectral range absorption when assembled into oriented moiré heterostructures[16,17]. Several works[18–20] also report on studies of the total optical response from graphene/$h$BN heterostructures. Nevertheless, the influence of $h$BN substrate or encapsulation on the intrinsic optical response of an almost transparent graphene[21,22] in the visible spectral range yet remains undetermined.

At the same time, the optical properties of graphene on standard substrates, such as SiO$_2$/Si, quartz and a variety of glasses were thoroughly investigated by spectroscopic ellipsometry[23–29]. Despite the non-identical fitting approaches and graphene samples (exfoliated or chemical vapor deposited), all works agree on the universal value of the absorption, which is defined by the fine-structure constant $α$. Nonetheless, several works[16,17,30] argue that this situation may change in the presence of $h$BN.

In this work, we present an experimental investigation of optical properties of graphene on $h$BN substrates through the imaging spectroscopic ellipsometry technique. We demonstrate an emergence of anomalous optical constants from monolayer graphene on top of a thick $h$BN and compare our results with the ones on one of the standard substrates (SiO$_2$/Si) from the literature and of our own. We also demonstrate a highly sensitive approach to the detailed analysis of ellipsometric parameters and optical response of graphene, which can potentially be easily extended to other two-dimensional materials.

**Results**

Before the optical measurements, we confirmed the quality of our exfoliated graphene samples on SiO$_2$/Si and $h$BN substrates by analyzing their structural properties. Figure 1 ((a) and (b)) display the schematics along with an optical image of one of our samples prepared on SiO$_2$/Si substrate through standard mechanical exfoliation technique. Acquired Raman spectrum suggests that it is a monolayer with a relative intensity ratio of *2D* to *G* peaks larger than 2 (see inset of Figure 1 (b)). Figure 1 (c) shows the results of rigorous examination of the surface morphology of our samples by atomic force microscopy (AFM). The roughness histogram of our SiO$_2$ substrates show a standard deviation of $σ_{SiO2}$ ~ 136 pm for a fitted Gaussian, which is slightly smaller, but in general consistent with the typical values reported elsewhere[5,31]. On the other hand, in the case of $h$BN substrates, we assemble another set of samples on transparent substrates (glass) through the dry transfer technique utilizing polycarbonate (PC) films[32,33]. Figure 1 ((d) and (e)) demonstrate the schematics and an optical image of one of the studied heterostructures. The inset of Figure 1 (e) demonstrates its Raman response proposing a composition of monolayer graphene with a thick $h$BN layer. Performed AFM scans reveal that our samples are free of nanoscale distortions or wrinkles of any kind (see Figure 1 (f)). A histogram of the roughness of the $h$BN layer shows a standard deviation of a fitted Gaussian measured to be $σ_{hBN}$ ~ 37 pm, which is about 3 times smaller than for SiO$_2$ substrates and is also consistent with the typical values reported elsewhere (see inset of Figure 1 (f))[5,31]. A prompt comparison of standard derivations measured for our graphene monolayers on SiO$_2$ and on hBN substrates confirms that those precisely nest on the surfaces of whatever they are placed on (see insets of Figure 1 ((c) and (f))). Thus, for the case of $h$BN substrates, we obtain ultraflat graphene layers with atomic smoothness.

We used imaging spectroscopic ellipsometry technique to characterize the optical response of our graphene monolayers on both substrates. The schematics of our experimental setup is shown in Figure 2 (a). The sensitivity of our technique allows us to study exfoliated and dry-transferred flakes in miniature regions of interest (10 μm$^2$) within the same field of view[34]. For both types of substrates, we performed a step-by-step analysis of the optical responses from the samples with and without graphene layers included (see Methods for further details). In the case of graphene on $h$BN substrate, the measured and calculated ellipsometric parameters $Ψ$ and $Δ$ are shown in Figure 2 ((b) and (c)), respectively. Those are in a good agreement as the ones for graphene on SiO$_2$/Si (see Supplementary Figure 1).

To obtain the dielectric function of graphene from the acquired ellipsometric spectra, we used the Drude-Lorentz oscillators model (see Methods), which considers the optical response of quasi-free electrons (Drude oscillator), and graphene's van Hove singularity for $π$-to-$π$* interband transitions (Lorentz oscillator)[23].

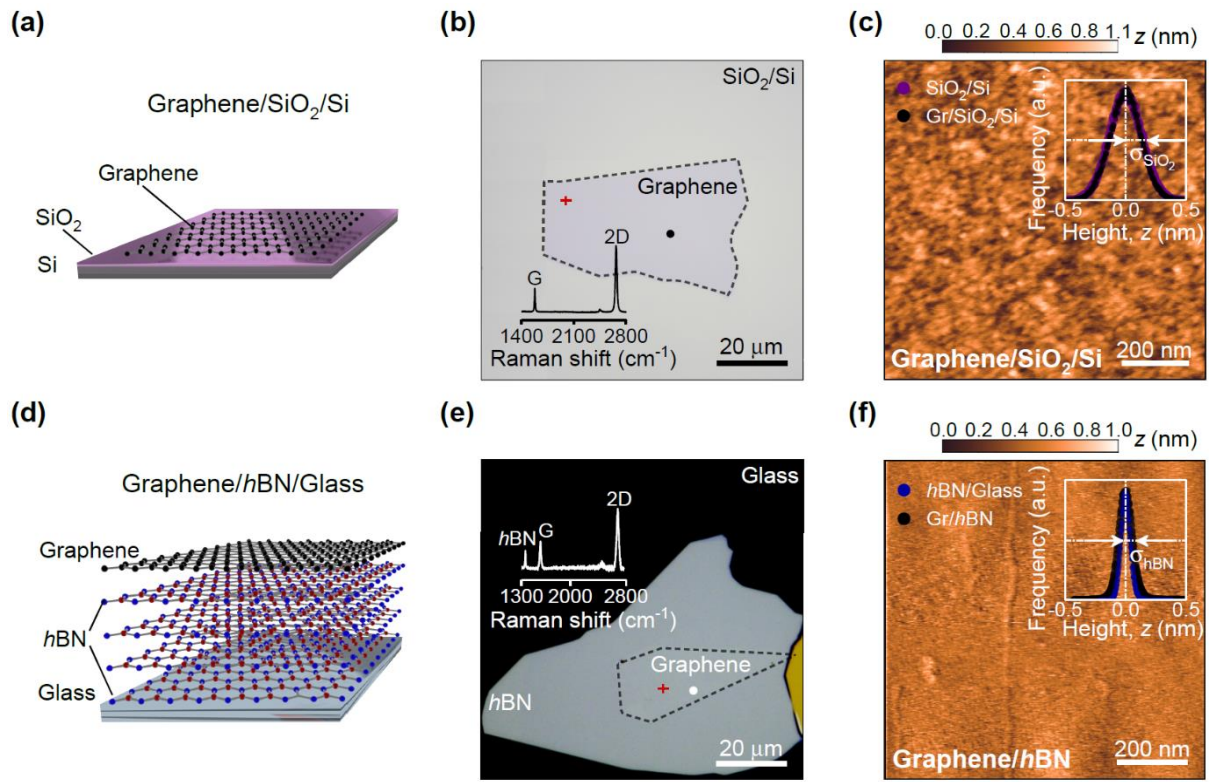

**Figure 1: Surface morphology of monolayer graphene on various substrates. (a)** Schematic illustration and **(b)** 50X optical image of graphene on SiO$_2$/Si substrate. Dashed lines are a guide to an eye emphasizing flake boundaries. Inset demonstrates obtained Raman spectrum taken from the point specified by black dot in (b). **(c)** Atomic force microscopy (AFM) colour map taken from the region specified by the red cross in (b). The colour bar shows the surface roughness. Inset shows histograms of the height distribution (surface roughness) for the substrate (SiO$_2$/Si) and the flake (graphene). **(d)** Schematic illustration and **(e)** 50X optical image of graphene on hBN/glass substrate. Dashed lines are a guide to an eye emphasizing flake boundaries. Inset demonstrates obtained Raman spectrum taken from the point specified by white dot in (e). **(f)** AFM colour map taken from the region specified by the red cross in (e). The colour bar shows the surface roughness. Inset shows histograms of the height distribution (surface roughness) for the substrate (hBN/glass) and the flake (graphene).

The determined real $Re[\varepsilon]$ and imaginary $Im[\varepsilon]$ parts of the dielectric function of graphene on both substrates are shown in Figure 2 ((d) and (e)), respectively. Unexpectedly, both parts of dielectric function of our ultraflat graphene on hBN (sample 1) are noticeably higher than for SiO$_2$ substrate (sample 2) in the whole interval of measured wavelengths. Additional investigations demonstrate a good repeatability (sample 3). To further validate our findings, we evaluated the transmittance spectrum of graphene/hBN heterostructure accounting for the acquired anomalous optical response of graphene and compared it to the measured one. Despite the observed excessive values of graphene's dielectric function, the theoretical transmittance spectrum matches well with the experimentally observed one as it can be seen in Figure 2 (f). In addition, to rule out the possibility that ambiguity of the used optical constants of hBN could have caused such an increment in graphene's optical response, we separately verified the optical response from the thick hBN flake using the same micro-transmittance technique (see Supplementary Figure 2).

In general, the optical responses from atomically thin layers are responsible for a very limited contribution compared to the substrate in the acquired ellipsometric spectra, which causes the accuracy of the measurements to fall. This comes to nearly an extreme case for monolayer graphene.

To enhance the sensitivity of our spectroscopic imaging technique, we assembled a specific configuration of layers giving a rise to a larger difference in optical responses from the substrate with and without graphene layer, and thus, to a higher sensitivity of ellipsometric parameters to graphene optical constants.

This is achieved in the vicinity of topological phase singularities, which arise owing to intersection of graphene optical constant's dispersion with the substrate zero-reflection surface[35]. Here, we dry-transferred another graphene/hBN heterostructure on top of a thick 200 nm Au film to ensure an appropriate formation of a cavity, shown in the schematics and the optical image in Figure 3 ((a) and (b)), for realization of topological phase singularity in the vicinity of ellipsometer's best sensitivity (around ~ 500 nm). The thickness of our hBN flake is 152 nm, which leads to a topological phase singularity at a wavelength of 477 nm. The corresponding ellipsometric parameter maps are presented in Figure 3 ((c) and (d)).

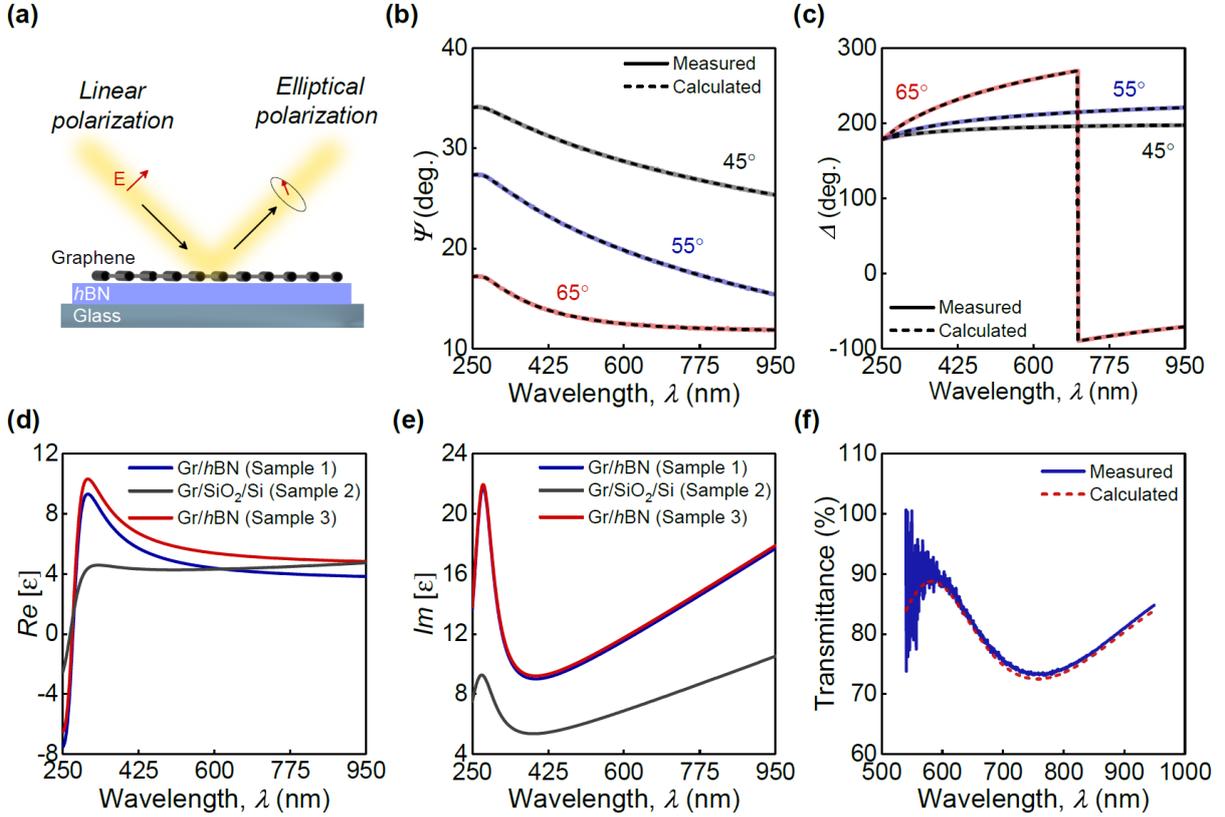

**Figure 2:** Imaging spectroscopic ellipsometry of monolayer graphene on hBN/glass substrate. **(a)** Schematic illustration of the measurement setup. Ellipsometric parameters $\Psi$ **(b)** and $\Delta$ **(c)** at three incident angles 45°, 55°, 65°. Solid (dashed) lines represent the measured (calculated) cases. Real **(d)** and imaginary **(e)** parts of the obtained dielectric function. Grey lines correspond to real and imaginary parts of dielectric function obtained for monolayer graphene on SiO$_2$/Si substrate. Drude-Lorentz oscillators parameters are collected in Supplementary Table 1. **(f)** Micro-transmittance spectra of graphene on hBN/glass substrate. Solid (dashed) line represents the measured (evaluated) case.

As expected, Figure 3 (d) shows a noticeable difference between graphene/hBN/Au and hBN/Au structures in $\Delta$ owing to constructed phase topology. As a result, at this point we have an increased sensitivity to graphene's optical response. Hence, this allows us to make a unique fit of the optical absorption of our graphene. Here, we calculated the difference between calculated and measured ellipsometry spectra with respect to graphene absorption in terms of mean squared error (MSE). Figure 3 (e) shows the resulting dependence of MSE of our measurements. Surprisingly, it reaches a minimum at values that are larger than $\pi\alpha$, where $\alpha$ is the fine structure constant[21,22], validating our high dielectric permittivity of graphene on hBN presented in Figure 2 (d) and (e). This suggests that the typical values of absorption could therefore be mended for our ultraflat graphene on hBN substrate.

The corresponding dependences of measured ellipsometric parameters on the wavelength in the vicinity of our topological phase singularity are demonstrated in Figure 3 ((f) and (g)) for $\Psi$ and $\Delta$, respectively. Insets demonstrate the apparent variation between the parameters evaluated for monolayer graphene accounting for two types of substrates studied. Notably, the variation is evidently smaller for the case of a graphene layer placed on an hBN substrate.

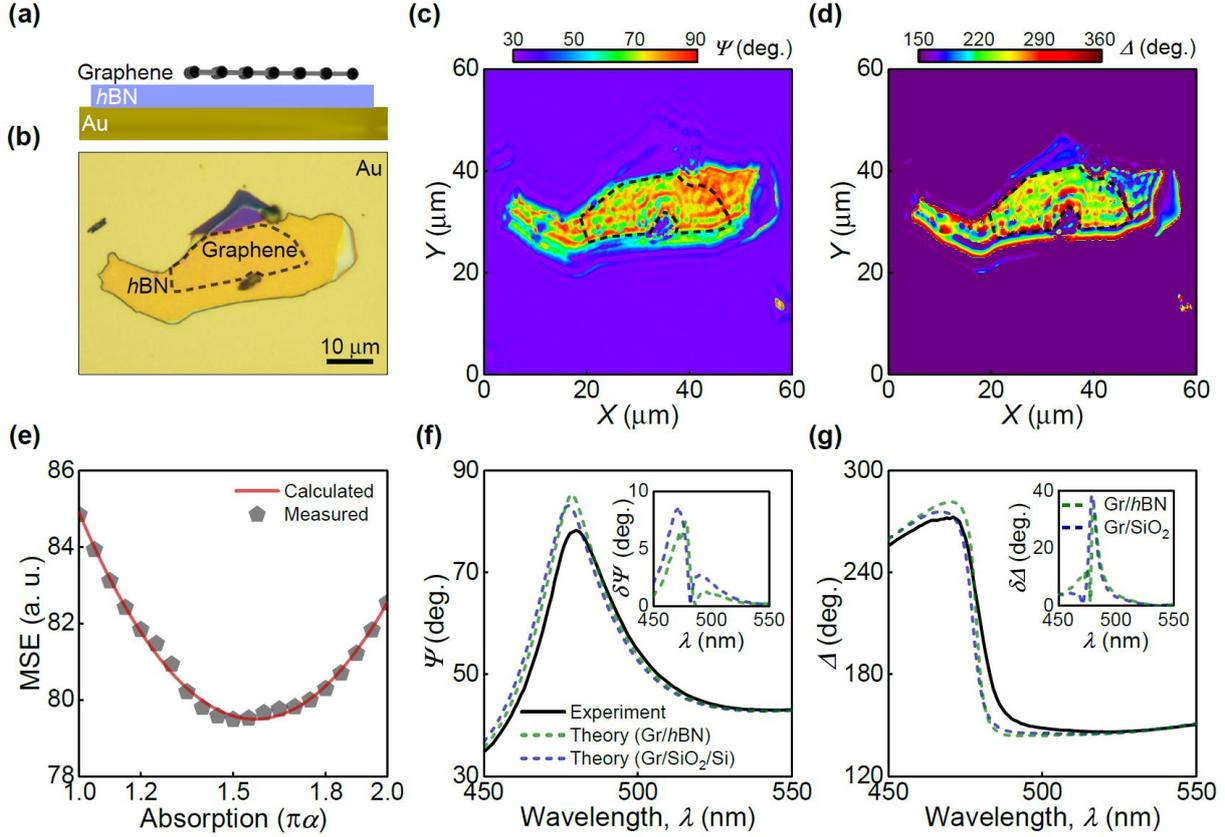

**Figure 3: Highly sensitive optical response to a monolayer graphene in the vicinity of topological phase singularity. (a)** Schematic illustration and **(b)** 50X optical image of graphene on hBN/Au substrate. Ellipsometric parameter colour maps of $\Psi$ **(c)** and $\Delta$ **(d)** at the wavelengths of 477 nm and 50° angle of incidence near the topological phase singularity. The colour bars show amplitude (c) and phase (d) distribution. Dashed lines are a guide to an eye emphasizing flake boundaries. **(e)** Evaluated (solid line) and measured (grey pentagons) mean squared error (MSE) dependence on the absorbance of graphene. Ellipsometric parameters $\Psi$ **(f)** and $\Delta$ **(g)** near the topological phase singularity. Solid lines represent the measured parameters for graphene on hBN/Au substrate. Dashed lines correspond to the evaluated parameters. Insets show the exact variation between the evaluated and the measured parameters.

Figure 4 demonstrates acquired dependencies of refractive indices, extinction coefficients, and the intrinsic absorbance of our exfoliated graphene samples on both types of substrates ($SiO_2$/Si and hBN) in comparison with literature data[24,26,27] (exfoliated graphene optical constants on a standard $SiO_2$/Si substrate). Despite the non-identical fitting approaches, all works report on universal optical responses for the case of graphene on $SiO_2$/Si, including our measurements.

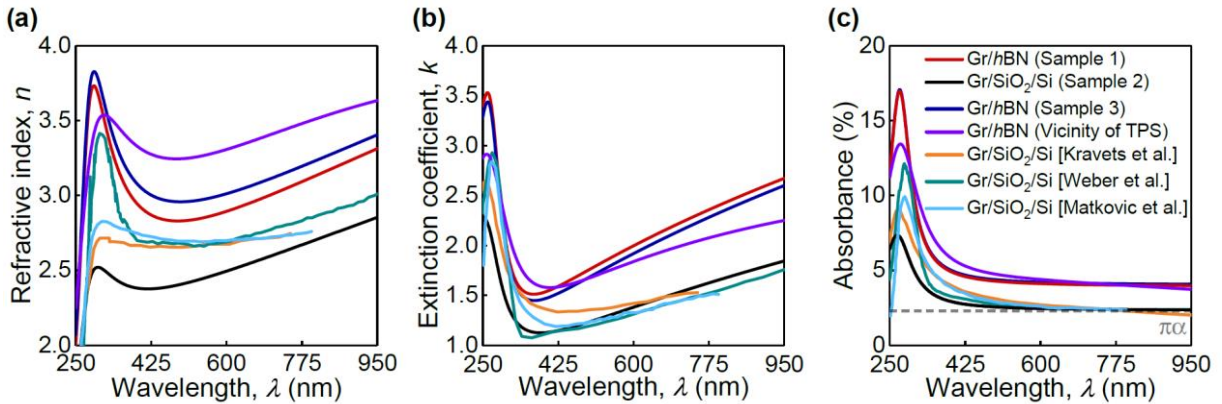

**Figure 4: Comparison of optical constants of exfoliated monolayer graphene on $SiO_2$/Si and hBN substrates. (a)** Refractive indices, **(b)** extinction coefficients and **(c)** intrinsic absorbance $A$ vs wavelength. $A = 4\pi nkt/\lambda$, where $n$ is refractive index, $k$ is extinction coefficient, $t$ is the thickness of graphene, and $\lambda$ is the wavelength of light.

On the other hand, graphene on *h*BN demonstrates substantially higher optical constants (Figure 4 (a) and (b)), compared to graphene on SiO$_2$/Si. For instance, graphene's refractive index and extinction coefficient is about 20 % and 40 % higher on *h*BN than on SiO$_2$/Si, which may be of use for the enhancement of absorption in graphene-based photonic devices[36,37]. In the case of an excitonic peak at 270 nm, the obtained behaviour can be explained by significant difference in static dielectric permittivities of SiO$_2$ ($\varepsilon_{SiO2}$ ~ 3.8) and *h*BN ($\varepsilon_{hBN}$ ~ 7)[38], which strongly affects excitonic optical response[39]. However, the situation in near-infrared range is more complicated since even high doping of graphene[40] should not affect its absorption in these spectral intervals (see Supplementary Note 1).

Nevertheless, angle-resolved photoemission spectroscopy studies[30] show that substrates with high dielectric permittivities can substantially modify the fine structure constants owing to emergence of electron-electron interactions. Indeed, our *ab-initio* calculations suggest that even a slight change in the interlayer distance between graphene and *h*BN may significantly affect the total optical response (see Supplementary Note 2). Other approaches also suggest notable growth of optical constants in graphene/van der Waals material heterostructures[41–44]. Nevertheless, further research is required to explain the physical mechanisms of such an increase in graphene's absorption when placed on top of *h*BN substrate, and other van der Waals materials.

**Discussion**

Integration of *h*BN and graphene into van der Waals heterostructures results in emergence of extraordinary electronic properties. Therefore, it is of fundamental and practical interest to study the influence of *h*BN on graphene's optical properties. Our imaging spectroscopic ellipsometry measurements showed that *h*BN substrates could substantially enhance the absorption in graphene by about 60 % in the broad spectral range (250—950 nm). Hence, those are more suitable than standard SiO$_2$/Si substrates for a variety of photonic applications, where the absorption plays a key role, such as photo- detection, modulation, and sensing. We attribute this behaviour to electron-electron interactions arising due to high static dielectric response of *h*BN. From a broader perspective, our studies reveal that the universal optical absorption of bare and pristine graphene can be reconstructed in the dielectric environment.

**Methods**

**Sample preparation.** We performed O$_2$ plasma-cleaning for all the types of substrates to enhance the adhesion with two-dimensional layers prior to exfoliation. Next, the substrates were heated up to 120 °C and the standard mechanical exfoliation from bulk graphite and *h*BN crystals was performed using commercial scotch tapes from "Nitto Denko Corporation". To integrate graphene monolayers with *h*BN, we used polymer based modified dry-transfer technique[32,33] established on utilization of double thin films; polydimethylsiloxane (PDMS) and polycarbonate (PC).

**Imaging spectroscopic ellipsometry.** To analyze optical constants of graphene samples, we used commercial imaging spectroscopic ellipsometer Accurion nanofilm_ep4 in the nulling operational mode. In our imaging ellipsometer, the spot size is about 2 mm in diameter. The high resolution is achieved not by focusing the light as it is usually done in classical ellipsometers, but by recording the image on a camera as it is shown in Figure 3(b) and (c). As a result, here, each of the pixels record the ellipsometric parameters, and allow us to take into account only the pixels that correspond to our sample. To avoid backside reflections, we used beam cutter following the approach presented by Funke and colleagues[45]. Ellipsometry spectra were recorded for the spectral range from ultraviolet (250 nm) to near-infrared (950 nm) for the samples on both types of substrates. During the measurements, we simultaneously recorded the ellipsometric signals from bare substrate and substrate with graphene. It allows us to determine the precise optical model of substrate to eliminate errors arising from slight inconsistency between literature optical constants and real one for substrate material (Si, SiO$_2$, glass, and *h*BN). Note that for the individual ellipsometric parameter analysis of our *h*BN substrates, we followed the algorithm described in Supplementary Note 2 of our recent work[34]. Afterwards, we fitted graphene optical constants with Drude-Lorentz optical model[23]:

$$\varepsilon(E) = \varepsilon_{1\infty} + \varepsilon_{Drude} + \varepsilon_{Lorentz} = \varepsilon_{1\infty} - \frac{\hbar^2}{\varepsilon_0 \rho (\tau E^2 + i\hbar E)} + \frac{ABE_0}{E_0^2 - E^2 - iBE}, \quad (1)$$

where $\varepsilon$ is the dielectric permittivity of graphene, $E$ is the photon energy in eV, $\varepsilon_{1\infty}$ is the offset of real part of dielectric permittivity, which takes into account absorption peaks for higher than measured energy range, $\hbar$ is the reduced Planck's constant, $\varepsilon_0$ is the vacuum dielectric constant, $\rho$ is the resistivity in $\Omega \cdot$cm, $\tau$ is the scattering time in sec., $A$ is the Lorentz oscillator strength, $B$ is the Lorentz broadening parameter, and $E_0$ is the Lorentz peak central energy. In accordance with AFM microscopy results (see Figure 1(c) and (f)), which show negligible roughness, we do not account for the roughness of our samples.

**Atomic force microscopy.** The morphology of all our samples was examined by AFM (NT-MDT Ntegra II). All measurements were performed in a dry state at room temperature using HybriD mode. AFM images were acquired using silicon tips (ScanSens, ETALON, HA_NC) with an elastic constant of 3.5 N/m and a resonance frequency of 140 kHz. The areas of 1 µm$^2$ with 400 pixels per line were obtained at a scanning rate of 0.2 Hz for all samples. The surface height distributions were extracted from areas of 0.2 µm$^2$ using Gwyddion software.

**Supplementary Information**

Supplementary information contains details on ellipsometric parameters of graphene on SiO$_2$/Si substrates, micro-transmittance spectra of bare *h*BN flakes, *ab-initio* calculations of optical properties of graphene on *h*BN substrates, and discussion of the influence of doping effect in graphene on its optical absorption.

**Acknowledgments**

This work was supported by Russian Science Foundation project No. 21-79-00218 and Ministry of Science and Higher Education of the Russian Federation Agreement No. 075-15-2022-1150 (topological phase singularity studies). The *ab-initio* calculations were performed using resources provided by to the Joint Supercomputer Center of the Russian Academy of Sciences.